\documentclass[magazine,draftcls,onecolumn]{IEEEtran}
\usepackage{bbm}
\usepackage{}
\usepackage{subfigure}
\usepackage[dvips]{graphicx}
\usepackage{graphicx}
\usepackage{pgf}
\usepackage{caption}
\usepackage{pifont}
\usepackage{bbding}
\usepackage{times}
\usepackage{cite}
\usepackage[numbers,sort&compress]{natbib}
\usepackage{times}
\usepackage{amsmath}
\usepackage{amssymb}
\usepackage{bm}
\usepackage{booktabs}
\usepackage{algorithmic}
\usepackage[lined,algonl,ruled]{algorithm2e}
\usepackage{multirow}
\usepackage{multicol,pdflscape}
\usepackage{url}
\usepackage{moreverb}

\newcommand\rb[1]{\raisebox{3ex}[0pt]{#1}}

\makeatletter
\def\ps@headings{%
\def\@oddhead{\mbox{}\scriptsize\rightmark \hfil \thepage}%
\def\@evenhead{\scriptsize\thepage \hfil \leftmark\mbox{}}%
\def\@oddfoot{}%
\def\@evenfoot{}}
\makeatother

\ifCLASSINFOpdf

\else

\fi

\begin{document}
\title{Connected Vehicular Transportation: Data Analytics and Traffic-dependent Networking}
\author{Cailian~Chen, Shanghai Jiao Tong University\\
\hspace{-1.7cm} Tom~Hao~Luan, Deakin University\\
\hspace{0.1cm} Xinping~Guan, Shanghai Jiao Tong University\\
\hspace{-1cm} Ning Lu, Thompson Rivers University\\
\hspace{-0.1cm}Yunshu~Liu, Shanghai Jiao Tong University
}
\maketitle

\begin{abstract}
With onboard operating systems becoming increasingly common in vehicles, the real-time broadband infotainment and Intelligent Transportation System (ITS) service applications in fast-motion vehicles become ever demanding, which are highly expected to significantly improve the efficiency and safety of our daily on-road lives. The emerging ITS and vehicular applications, e.g., trip planning, however, require substantial efforts on the real-time pervasive information collection and big data processing so as to provide quick decision making and feedbacks to the fast moving vehicles, which thus impose the significant challenges on the development of an efficient vehicular communication platform. In this article, we present TrasoNET, an integrated network framework to provide realtime intelligent transportation services to connected vehicles by exploring the data analytics and networking techniques. TrasoNET is built upon two key components. The first one guides vehicles to the appropriate access networks by exploring the information of realtime traffic status, specific user preferences, service applications and network conditions. The second component mainly involves a distributed automatic access engine, which enables individual vehicles to make distributed access decisions based on access recommender, local observation and historic information. We showcase the application of TrasoNET in a case study on real-time traffic sensing based on real traces of taxis.
\end{abstract}

\section{Introduction}
Our earth is facing the unstoppable increasing trend of vehicles. In United States, there are on average $812$ cars for every $1,000$ people. In China, the amount of vehicles is estimated to be $250$ million in 2020. The massive increase in vehicles has brought a series of social and environmental issues to our cities and daily lives such as frequent traffic jams, vehicle crashes, throat-choking air pollution, etc. A sustainable, intelligent and green transportation system is thus of crucial importance. Towards this goal, one practical solution is to use the cutting-edge wireless information and communication technologies to provide real-time transport-related information services to road administrators and vehicles, namely Connected Vehicular Transportation System (CVTS) \cite{MConti2014VTM,KAbboud2014ITS,WHZhuang2011Infotainment}. As a result, the transportation efficiency can be significantly improved with more smooth traffic flows and travellers can get informed for more wise route selections and enhanced travel experience. Furthermore, both Google and Apple released their mobile operating systems for autos in 2014. It is estimated that the global Connected Car Market will reach $30.2$ billion in 2015, and $80\%$ of all autos sold in 2016 will be connected. Therefore, it is foreseeable that in near future, connected vehicles would embody a pragmatic solution towards Intelligent Transportation System (ITS).

CVTS aims to make safer and more coordinated use of transportation networks. Such a system lies on the timely collection of road traffic information, effective data analytics, and quick decision making and feedbacks to the traffic management facilities and vehicles. Besides the roadside sensors (e.g. GPS, cameras, inductive loops, RFID and in-road reflectors) deployed in the city-wide for traditional traffic sensing, connected vehicles provide a new efficient traffic monitoring method by the live data streams of large number of the off-the-shelf mobile terminals (e.g., on-board wireless communication facilities, smartphones, tachographs and wearable devices). It is envisioned that with the adoptions of the embedded, tethered or smartphone integrated vehicular sensing and communication facilities, the demand of on board infotainment services would become more demanding, which eventually would generate a large volume of data required for processing. Moreover, the large variety of data sources and applications require CVTS with the assets of fast response and processing rate, and more importantly, with high accuracy, reliability and security.

With the increasingly growing data in CVTS \cite{JFWan2014VCPS}, there rise the fundamental engineering challenges from the following three aspects: (i) big data collection from ubiquitous roadside and in-vehicle sensors in the city; (ii) deep data analysis in traffic management center; and (iii) real-time decisions returned to traffic management facilities and vehicles. In this cycle, the first step is to timely, effectively and economically collect monitoring data from the ubiquitous sensors in the city. Compared to the dramatic improvement on technical tools for handling data \cite{MGramaglia2014ABEONA}, vehicular networking for CVTS applications has, however, adapted much slower towards the low-cost and efficient data collection, which motives this article.

In this article, we unfold our journey by first reviewing the impact of data analysis on some representative realtime traffic-related services. The basic requirements for vehicular networking architecture are then identified. It is followed by a traffic-dependent network architecture for traffic data collecting and efficient service provisioning. Lastly, a case study is presented to show how realtime traffic estimation and timely network access can be implemented under the proposed architecture. The main contributions of this paper are summarized as follows:
\begin{itemize}
\item{A novel Traffic-Social Network framework, called TrasoNET, is presented for the first time to build the connection of realtime traffic and networking. Under this framework, the data analytics of CVTS take effects on the macroscopic, midscopic and microscopic network resource allocation and network access. Networking information is an effective data source for traffic sensing. It makes data analytics and traffic-dependent networking mutually beneficial, which is the core idea of this work.}
\item{A new traffic-dependent network access scheme is developed with network access recommendation from higher layer (network) and distributed automatic access decision-making in lower layer (terminal). It enables individual vehicles to make access decisions based on access recommender and local observation on network conditions.}
\item{A case study is presented to show the real data analytics for traffic estimation in Shanghai, China. Extensive simulations demonstrate that TrasoNEt can effectively select optimum network to ensure QoS of vehicles/ mobile devices, and network resource is fully utilized without network congestions by data offloading.}
\end{itemize}

\section{Data Analytics for CVTS Applications}   \label{sec:bigdatainITS}
The continuous monitoring on movements (e.g., safety services by short range v2x communications), mobility applications and vehicles condition monitoring would result in exponential growth of diverse source data which provides a wealth of information that is valuable to traffic management parties, drivers, repair shops and automakers.

In the following, we list some representative CVTS applications to describe their dependance on data analytics.

\begin{figure}[t]
\centering
\includegraphics[width=4.5in]{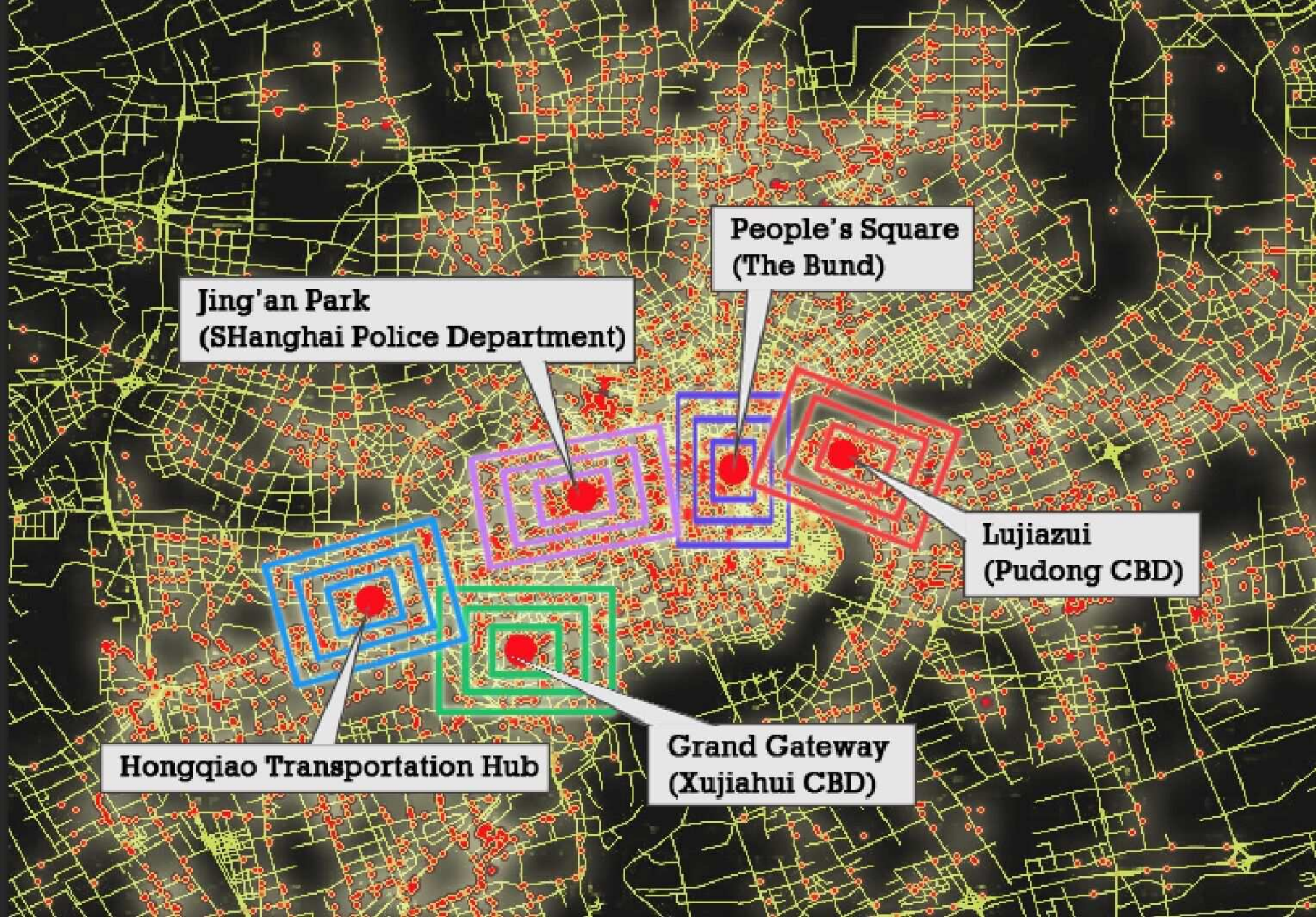}
\caption{{The spacial-temporal average traffic density based on GPS dataset of taxies in Shanghai, China on Jan. 24, 2013.}}
  \label{fig:GPSdata}
  \vspace{-0.3cm}
\end{figure}

\subsubsection{Real-time traffic estimation}
By using the moving vehicles or smartphones on-board vehicles for data sensing, and uploading the sensing reports (such as time, location and heading direction of vehicles) to the data center, the realtime traffic conditions of the roads, such as average running speed and traffic density, can be achieved by data analysis.

The taxi/bus management system in Shanghai, China, represents a practical deployment of the CVTS platform. Around $40,000$ taxis and buses of Shanghai are now equipped with the on-board GPS and sensors which periodically report the vehicle information (GPS location, velocity, heading direction, passengers on/off) in cycles ranging from $30$ seconds to $5$ minutes. This results in $65$ million records transmitted to the traffic management center everyday through cellular networks, and enables multiple management purposes. Road traffic conditions can be estimated efficiently by sparse sensing and advanced estimation methods. For example, compressive sensing and matrix completion based methods are reported in \cite{RDu2015,RDuGlobecom2013} based on the GPS dataset of Shanghai.

\subsubsection{Online navigation for connected vehicles}
Traffic prediction is more difficult than traffic estimation. Fortunately, through correlation analysis of big data, traffic patterns can be gleaned more easily, faster and clearly than before. For example, a social proximity mobility pattern of vehicles is adopted in \cite{NLu2014springer}, i.e., each vehicle has a restricted mobility region around a specific social spot such as a financial and sport center. By using data analysis, the social spots of vehicles in the real-world can be identified. The traffic peak probably appear at the rush hour around the social spot (SP), which makes traffics predictable. Five SPs has been shown in Fig. \ref{fig:GPSdata}. In another example, the traffic can also be predicted from the message published in social networks, such as network group events (e.g., big show and football game) information including time, place, and number of attendees. The traffic can be predicted to influx to the social places before the event and outflow after the event.

The traffic also has strong correlations with online navigation services: the more people in a particular geographic place search for the routes to a particular destination online, the more probably the traffic congestion happens on the route to the destination. In this case, the traffic can be predicted with the data from search engine. The online navigation server can provide more feasible path plans for vehicles.

\subsubsection{Remote vehicle diagnostics and road condition warning}
Based on the data collected from cars, the drivers can arrange better service interval by taking their own driving habits, and predicted wear and tear into account rather than conventional means based on defined number of kilometers. More importantly, valuable information on potential vehicle or road hazards can be delivered to drivers in realtime. For example, if a number of vehicles' traction control systems are activated at the same time and place, the cars in this area can be warned about ``icy conditions", fuel-efficient driving in heavy traffic, and etc.

\subsubsection{Fuel up or charge}
During travelling in the city, the electric vehicles/hybrid electric vehicles (EVs/HEVs) may make decision on what time and where to fuel up or load at location-specific charging piles by estimating the driving mileage according to real-time traffic conditions. The information provided by CVTS is valuable to design efficient ways of resource management in smart grid system in the city \cite{MWang2014JSAC}.

\subsubsection{Dynamic urban planning}
Sensory data from vehicles and mobile devices provide a pervasive way to understand how people use the city's infrastructure and affect the city, including urban dynamics, energy consumption and environment impacts such as noise and pollution. The big data related to real-time traffic can be used to improve city's services. For example, the planning and management parties can estimate how residential and working areas in cities are connected temporally, what the dynamic correlation of traffic density and pollution level appears, and how to reduce operational costs by optimizing planning. It also creates feedback loops with vehicles to reduce energy consumption and environmental impact.

\section{Features of Data Analytics in CVTS}
For the aforementioned applications, to explore the strong correlation among multi-source data is the key, which helps us capture the present road conditions and predict the future. The data collection for connected vehicles and related applications of CVTS distinguish with the traditional ones from the following three aspects:
\begin{itemize}
\item{\emph{From static sensing to dynamic sensing}: As a large amount of traffic data can now be harnessed through ubiquitous roadside sensors, vehicles and mobile devices, static traffic sampling no longer makes as much sense. Moreover, due to the complicated traffic conditions, accurate traffic estimation and prediction can hardly be achieved on small random samples, and require as much data as possible. Connected vehicles make it possible.}
\item{\emph{From precise data to messy data} \cite{BigData2013}: In the applications, allowing for imprecision (for messiness) of data may be an advantage, rather than a shortcoming. We can infer the vehicles' direction, speed and position with messy GPS data thus traffic estimation can be improved to the level of predicting the traffic congestion in a particular road rather than a region in the city by using 65 millions of ``dirty" (or ``noisy") taxi GPS data rather than small precise samples from digital camera and loop detector \cite{RDu2015}. The data could be messy if they cover as many streets as possible.}
\item{\emph{From parametric data to nonparametric data}: Correlations are useful in a small-data world, but they really shine in the context of large volume data and/or big data.  For traffic sensing and prediction, different (complicated) traffic models can be assumed to parameterize the relationship of traffic flows. However, multi-source data in CVTS allows us to pick a nonparametric model with simpler algorithms, and it results in more accurate than the sophisticated solution \cite{RDu2015,BigData2013}.}
\end{itemize}

To summarize, for data analytics of connected vehicles, we can relax the standards of allowable errors and increase messiness by combining different types of data information from different sources. In dealing with even more comprehensive datasets, we no longer need to worry so much about individual data points, but biasing the overall analysis. Through them we can glean insights more easily, faster, and more clearly than before. Correlations of multi-source data let us analyze the city traffic not by shedding light on its inner working but by identifying a useful proxy (e.g., the online navigation requests implies the possible congestion) for it. It is foreseeable that big data enhances the data analytics in CVTS to change the way of services provisioning in and enables connected vehicles from multi-dimensions of traffic, vehicular network and ITS.

\begin{figure*}[!t]
  \centering
  \includegraphics[width=6in]{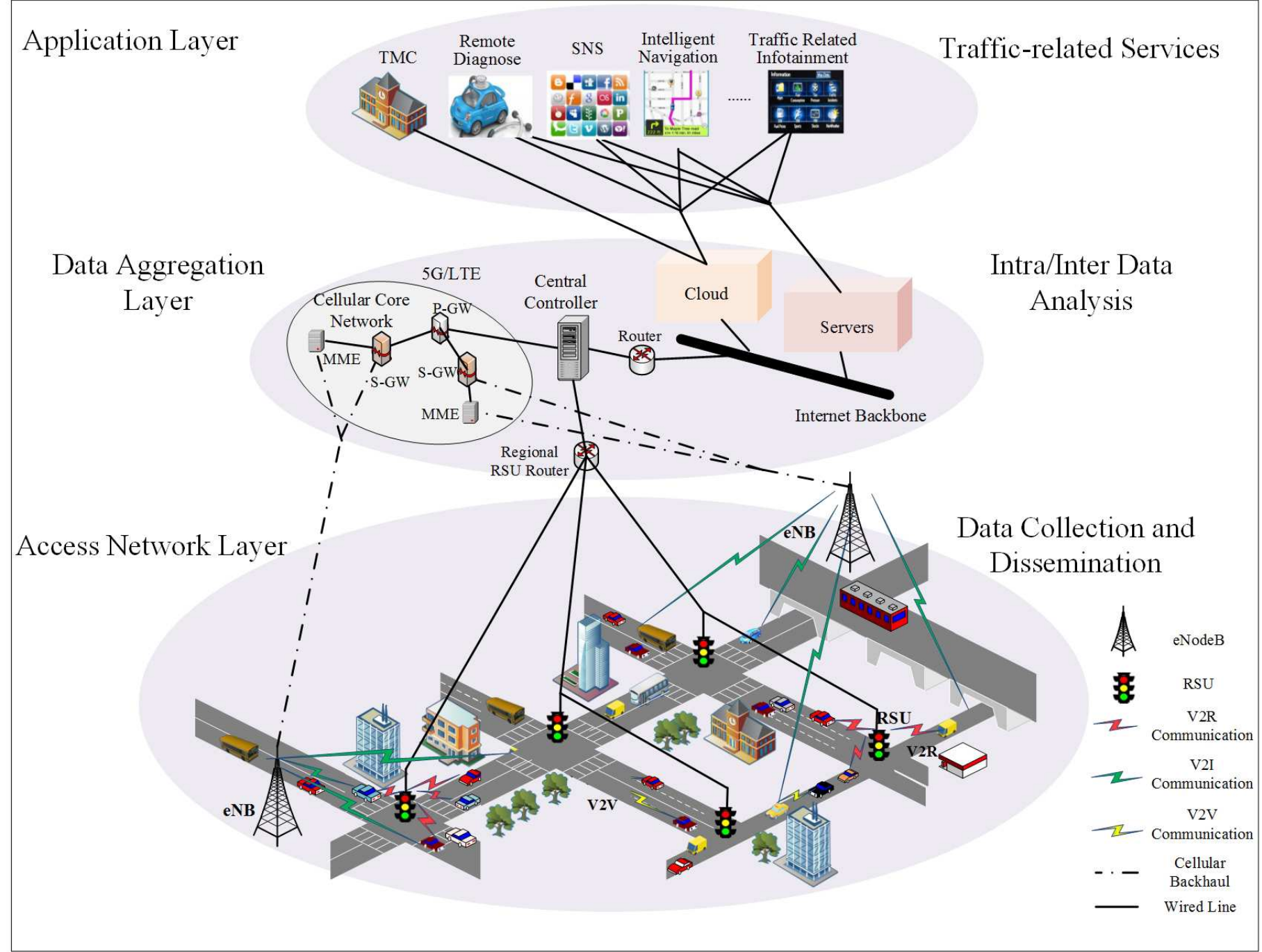}
  \caption{Network architecture for connected vehicles}
  \label{fig:arch}
\end{figure*}

\section{Traffic-dependent Networking for CVTS}

To engineer an efficient and economic network architecture is the foremost issue to facilitate the data collection, decision feedbacks and traffic related services.

\subsection{Network Framework}
The network framework is challenged by following issues:
\begin{itemize}
\item From a traffic sensing perspective, even with the broad mobility of vehicles and the dense deployment of static sensors on the road, it cannot be guaranteed that the traffic information for all the roads in all the time could be sensed. It therefore calls for an efficient and economical sampling way for traffic sensing. Crowdsensing by more vehicles could be one of the solutions in the very near future.
\item From a social perspective, the spatial distribution of traffic may follow a specific social pattern such as the power-law distribution features for the traffic in Shanghai, China which is shown in Fig.\ref{fig:GPSdata}. The traffic density decays from the business hot spots towards the boarder of the vehicles' mobility regions. One of the main challenges is finding a good guidance for wireless access to heterogeneous wireless networks (cellular network and vehicular ad hoc networks) such that the distributed traffic-dependent service requests can be satisfied with good Quality of Service (QoS)\cite{magazine_paper2}.
\end{itemize}

In this article, we describe a Traffic-Social Network (TrasoNET) framework for CVTS as in Fig.~\ref{fig:arch} to support the crowdsensing and network access guidance according to realtime traffic and traffic pattern.

TrasoNET consists of three layers: \emph{access network layer}, \emph{data aggregation layer} and \emph{application layer}. In the access network layer, sensory nodes, including vehicles and the mobile devices, could connect to roadside communication infrastructures (e.g., cellular base stations and roadside units) and communicate through LTE/5G cellular networks and/or vehicular ad hoc networks (VANETs). The static sensors (e.g. cameras, inductive loops, RFID and in-road reflectors) transmit data through wired communication. In the data aggregation layer, the roadside communication infrastructure are connected to corresponding backbone routers. Data flows are combined through the so-called \emph{central controller sub-layer} or so-called fog computing server, and further delivered to the cloud server through Internet. In the \emph{application layer}, the traffic management center (TMC) aggregates the collected multi-source data from cloud and analyze the data to estimate and predict the road traffic. The cloud also connects to other service providers such that the traffic-related information can be fused out and provided in the application layer. Different traffic-related services are then delivered to vehicles through cellular core network and regional VANET.

We elaborate on the four core components in the framework to highlight the characteristics in traffic-dependent networking.

\textbf{Infrastructure}: The access infrastructures consists of the evolved NodeBs (eNBs) and RSUs. It is assumed eNBs cover the whole city, and the communication link between mobile device and eNB is more stable than that between mobile device with RSU. RSU is equipped with a wireless transceiver operating on DSRC and/or WiFi, and hence the transmission range is small compared with eNB. But it provides high-rate transmission for mobile devices. Due to the explosive growth of mobile data traffic, the cellular network nowadays is straining to meet the current mobile data demand and faces an increasingly severe overload problem. RSU is not only an alternative for V2I (vehicular to infrastructure) communications, but also enables an offloading for cellular networks \cite{NLu2014springer}.

\textbf{Mobile Devices}: We do not discriminate what kind of mobile devices they are, but care about what network they access to. Normally, smartphones can connect to cellular network through LTE/5G and VANET infrastructures through WiFi, while vehicles can additionally connect to VANET infrastructure and other vehicles through DSRC. Since WiFi and DSRC technologies can be applied to drive-thru connection when they are moving on the road \cite{THLuan2012}, we propose an automatic network access engine in the mobile devices to offload data originally targeted for cellular networks, which is referred to as the automatic offloading engine.

\textbf{Central Controller}: The central controller is connected to base stations (e.g., eNBs for LTE), RSUs and Internet backbones. It allocates the network radio resources based on the realtime traffic estimated by TMC, and service demands requested by mobile devices. It acts as an interface between the physical network routers and the network operators to specify network services. The controller builds a logical control plane separated from data plane. Different from Internet Protocol (IP) based networks, such a frame enables mobile devices to move between different access interfaces without changing identities or violating specifications. The control function can be implemented by a protocol known as OpenFlow which enable controller to drive the access network edge hardware in order to create an easily programmable identity-based overlay on the traditional IP core.

\textbf{Cloud}: As the data analytics center for TMC and other service providers, the cloud receives data from static traffic sensors and mobile devices, and analyzes them for traffic estimation and prediction. Other traffic-related services are then analyzed based on the realtime traffic and data from other service providers. One key feature provided by cloud is the access guidance for the mobile devices to facilitate the automatic offloading engine.

\begin{figure*}[!t]
  \centering
  \includegraphics[width=6in]{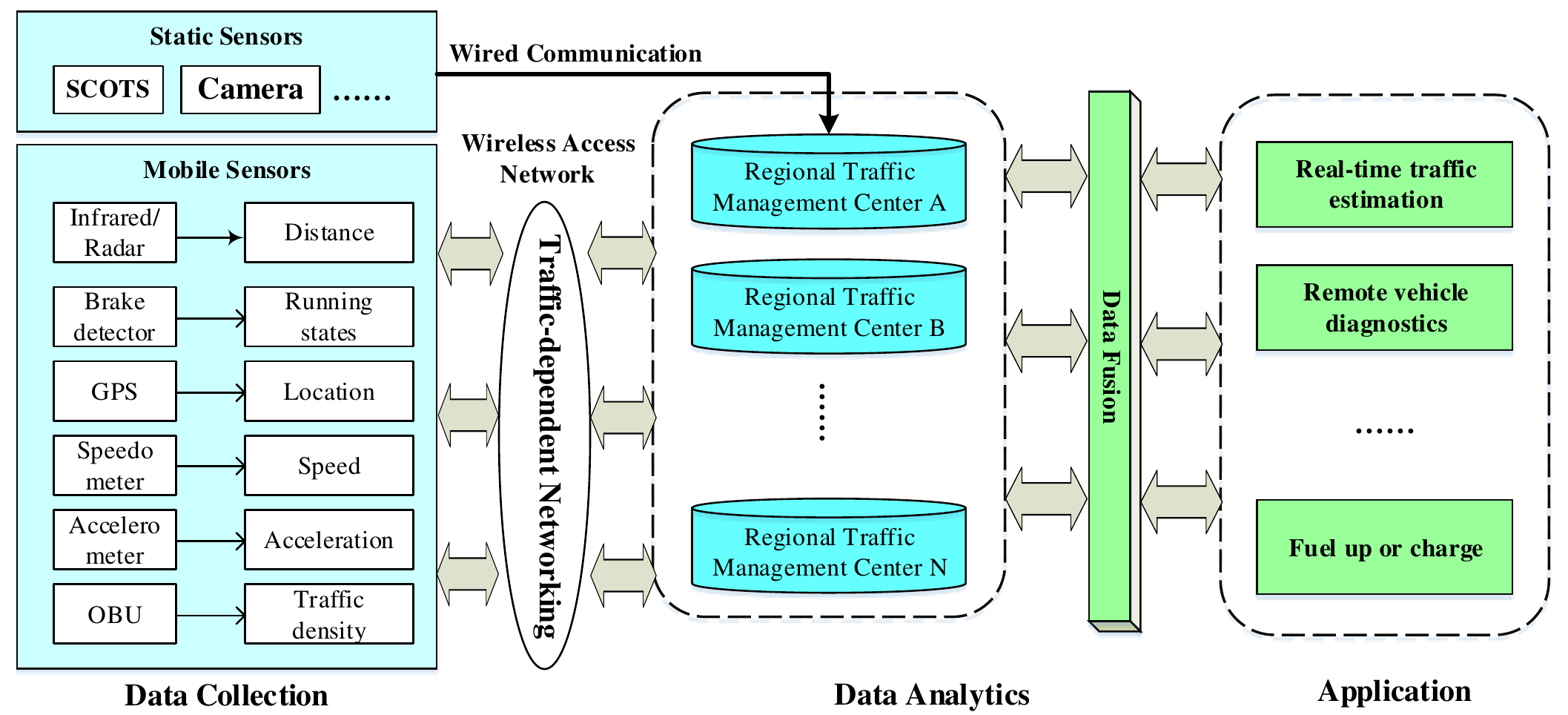}
  \caption{Data collection, analytics and applications}
  \label{fig:data}
\end{figure*}

As it is shown in Fig. \ref{fig:data}, TrasoNET builds the connection of data collection, analytics and traffic-dependent networking from the following three aspects:
\begin{enumerate}
\item{Firstly, the traffic big data are collected from static and mobile sensors through access network of TrasoNET. The aforementioned static sensors (e.g., cameras and inductive loops) transmit the traffic data to Regional Traffic Management Center through wired networks. Ubiquitous data from Mobile Devices (e.g., embedded, tethered or integrated on-board units, and smartphone) could be transmitted through wireless access network. For example, the probe vehicles (such as Taxis and buses) and floating cars (such as police cars from Public Security Bureau) in the city could provide sparse GPS data for preliminary traffic estimation. Then the traffic-dependent networking mechanism to be introduced in Subsection \ref{Access} could facilitate big data collection from ubiquitous Mobile Devices. More data improves the traffic estimation and other traffic related services.}

\item{Secondly, on the aggregation layer and application layer of TrasoNET, the data analytics provides the real-time regional and global traffic conditions. It facilitates the Central Controller to allocate wide-area network radio resources (e.g., base stations, RSUs and Internet backbones) according to the estimated traffic density, speed, acceleration and other information of vehicles/users in the city. Another key feature dependent on real-time traffic condition is the regional network access guidance for Mobile Devices, which realizes locally network resource management. As for Mobile Devices, the decision-making of network selection and handover can be given locally by the guidance-based access mechanism for efficiency and offloading purposes. In this sense, the data analytics take effects on the macroscopic, midscopic and microscopic network resource allocation and network access.}

\item{Thirdly, the various data from different network components provide complimentary data for deep data analytics. For example, the number of vehicles connected to an access point of wireless communication can reflect vehicle density, which can reduce the cost for satisfactory traffic estimation accuracy compared to traditional sensing methods with digital cameras and loop detectors. Besides the realtime estimation, the TrasoNET facilitates online navigation, remote vehicle diagnostics, fuel up and charge and other emerging applications.}
\end{enumerate}

\subsection{Traffic-dependent Network Access Mechanism} \label{Access}
The access control of networks is one of the key mechanism to guarantee the real-time CVTS applications. In the framework of TrasoNET, we give a guidance-based automatic access mechanism for efficient and offloading purposes. From the perspective of network access, the aforementioned four components map the phases of guidance, information push and distributed decision-making into Access Recommender Console, Broadcasting and Automatic offloading Engine.

\begin{figure}[tbp]
\centering
\includegraphics[width=6in]{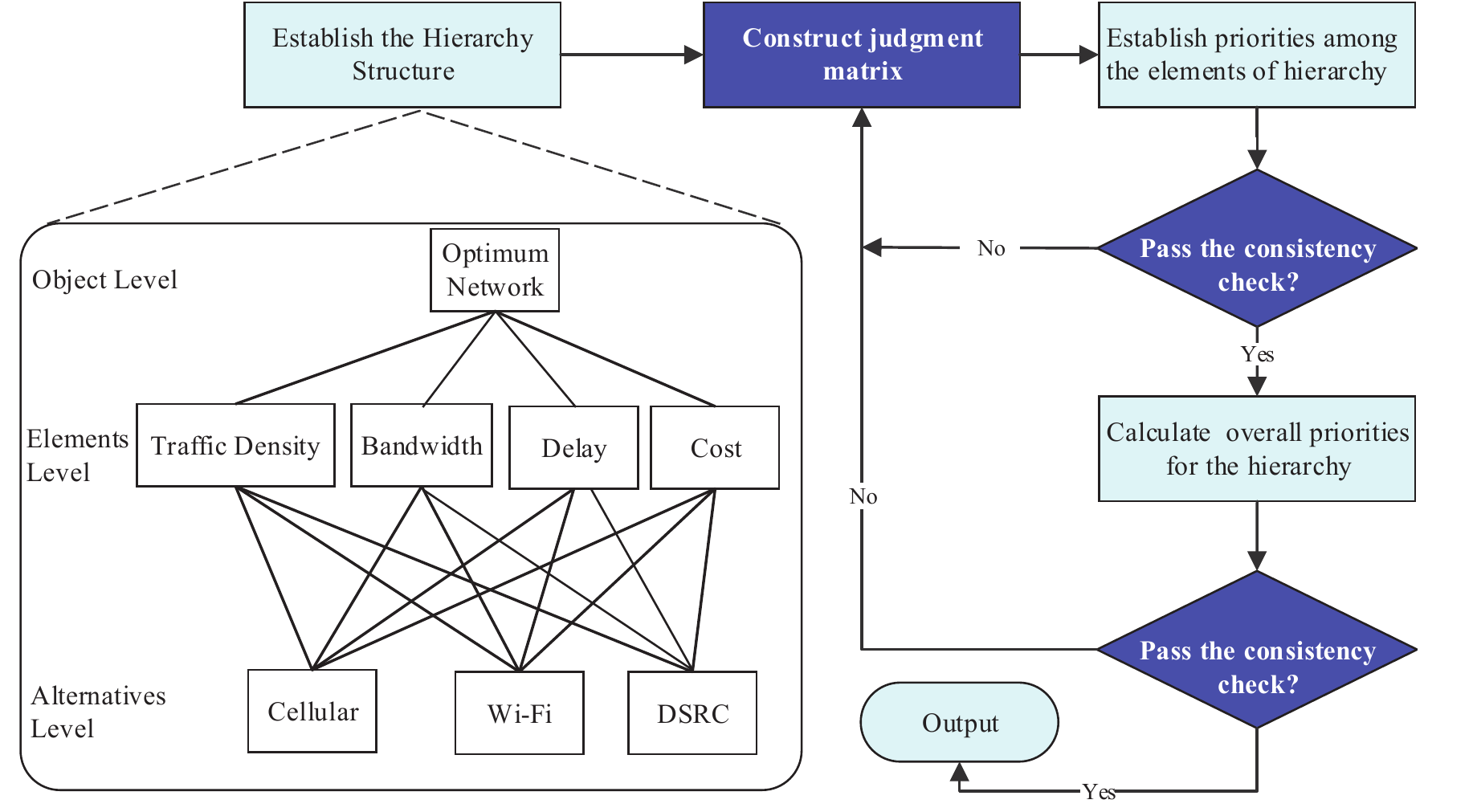}
\caption{Analytic Hierarchy Process for network selection}
\label{fig:AHPAlgorithm}
\end{figure}

\hspace{-.35cm}\textbf{Access recommender console}: To recommend an ``optimum network'' to vehicles based on multiple criteria, the cloud could apply intelligent computation methods to set the priority of network access for a specific region under realtime traffic condition. The well-known Analytic Hierarchy Process (AHP) for multi-criteria decision \cite{AHP2008} is one of good solutions. The logical flowchart of AHP algorithm is given in Fig. \ref{fig:AHPAlgorithm}. The key steps are introduced in the following.
\begin{itemize}
\item Model the network recommendation problem as a hierarchy which contains the goal, alternatives for reaching the goal, and criteria for evaluating alternatives.

\item Establish priorities among the elements of the hierarchy by making a series of judgments based on pair-wise comparisons of these elements. The compared results construct a pair-wise comparison matrix $A={a_ij},i,j=1,2,\cdots,n$, where $n$ is the number of criteria of second level,and every element $a_{ij}$ is based on a standardized comparison scale from \emph{equal importance} to \emph{dominance}.

\item The pair-wise comparison matrix should satisfy transitive preference and strength relations, it is necessary to check its consistency. Calculate consistency indicators $C.I.$, random consistency indicators $RI$, and get the consistency ratio $CR=CI/RI$. For example, consistency of judgement matrix is acceptable for the case of $CR<0.1$.

\item Synthesize these priority vectors to construct an overall priority vector and check the consistency again.
\end{itemize}

The final priorities of alternative networks for the ``Optimum Network'' can be got through the above algorithm. Traffic density is the critical factor and reflects the feature of vehicles' mobility.

\begin{figure}[tbp]
\centering
\includegraphics[width=6.5in]{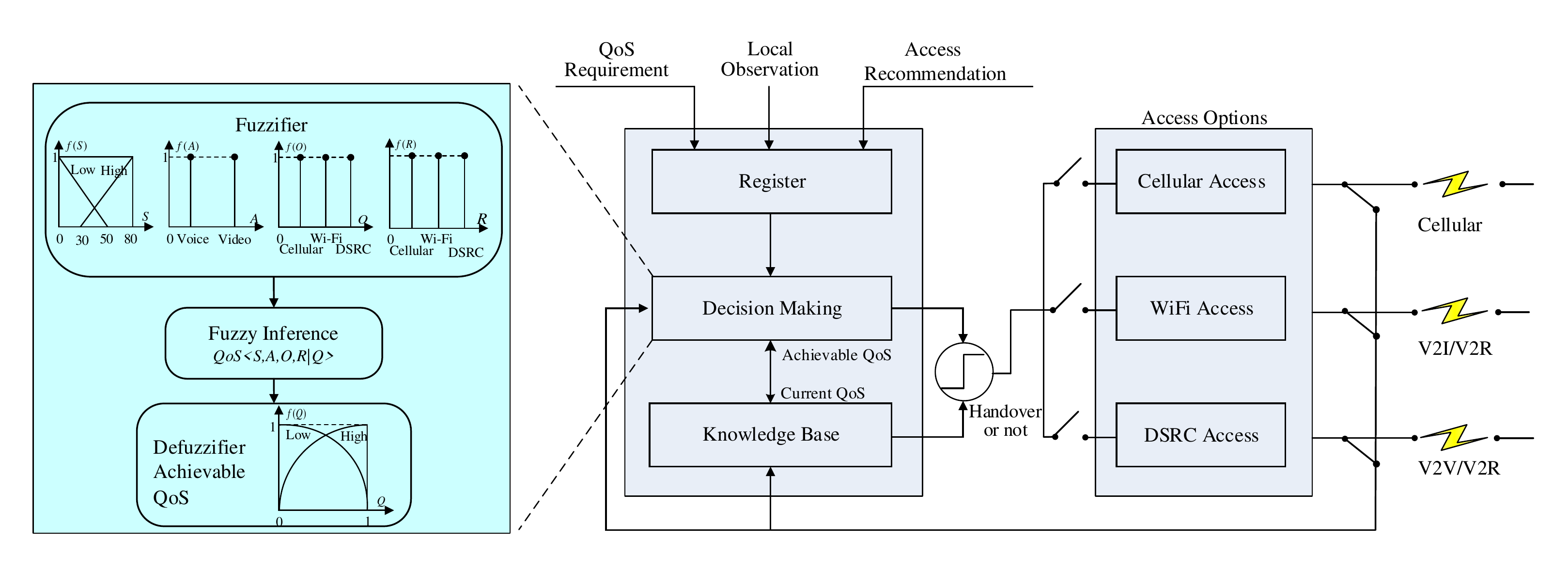}
\caption{Distributed decision making process}
\label{fig:app}
\end{figure}

\hspace{-.35cm}\textbf{Distributed Automatic Access Engine}: The engine operators in an automatic process shown in Fig. \ref{fig:app}. The QoS requirements ($\langle \textit{data rate, delay, cost} \rangle$) of various applications are registered with local observation of vehicle speed and the access recommender pushed through cellular network. The access option can be decided by analyzing the registered information, the received signal strength (RSS) of communication links and the statistical knowledge in the past. It is noted that the knowledge base is defined as
\begin{eqnarray*}
\mathcal{Q}\langle \textit{Speed, Application, Access option}, QoS|\textit{Access Recommender}\rangle,
\end{eqnarray*}
which can be abbreviated as $\mathcal{Q}\langle S,A,O,Q|R \rangle$. The knowledge base could be updated by the new achieved QoS periodically. The trustworthiness on access recommender can be adapted according to local observation and achieved QoS (access trials or QoS in a specific accessed network) for device's access decision-making (handover to another access network or not). The adapted process can be implemented by designing proper low-complexity algorithm in APP such as in Fig. \ref{SHscheme} through rule based inference \cite{CLChen2005TFS} for decision-making. In this process, the aforementioned proximity traffic pattern, locations of infrastructure and RSS statistics are the preferences for consideration so that the rules could be logically given.

In the following, fuzzy rules are powerful to represent the relation between the achieved QoS under accessed network and the criteria $\langle\textit{S,A,O,R}\rangle$ for automatic access engine. In fuzzy theory, a rulebase is a function $F$ that maps an input vector into outputs. Here, the premise variables are set as the four factors $\langle\textit{S,A,O,R}\rangle$. The achievable QoS level is defined as the output. The membership function for each variable can be defined. It could be simplified into singleton fuzzified levels for each premise variable. For example, set $Low$ and $High$ for $S$, classify $Voice, Text$ and $Video$ for $A$, and let $Cellular, WiFi$ and $VANET$ for both of access recommender $R$ and access option of the engine. An exemplary fuzzy rule with $l$ levels of output could be as follows:
\begin{description}
  \item[Rule $i$: ] \hspace{0.1cm}If $S$ is $Low$, $A$ is $Voice$, $O$ is $Cellular$ and $R$ is $Cellular$, then the achievable QoS could be $Level_l$.
\end{description}

Comparing the achievable QoS $Level_l$ through fuzzy decision-making and the achieved QoS $Level_c$, we can decide whether or not to handover to the ``optimum network". Only if achievable QoS $Level_l$ is better in a certain degree than the achieved QoS $Level_c$, the handover happens.

\section{Traffic sensing and traffic-dependent networking: A Case Study}
\begin{figure}[htbp]
\centering
\includegraphics[width=5in]{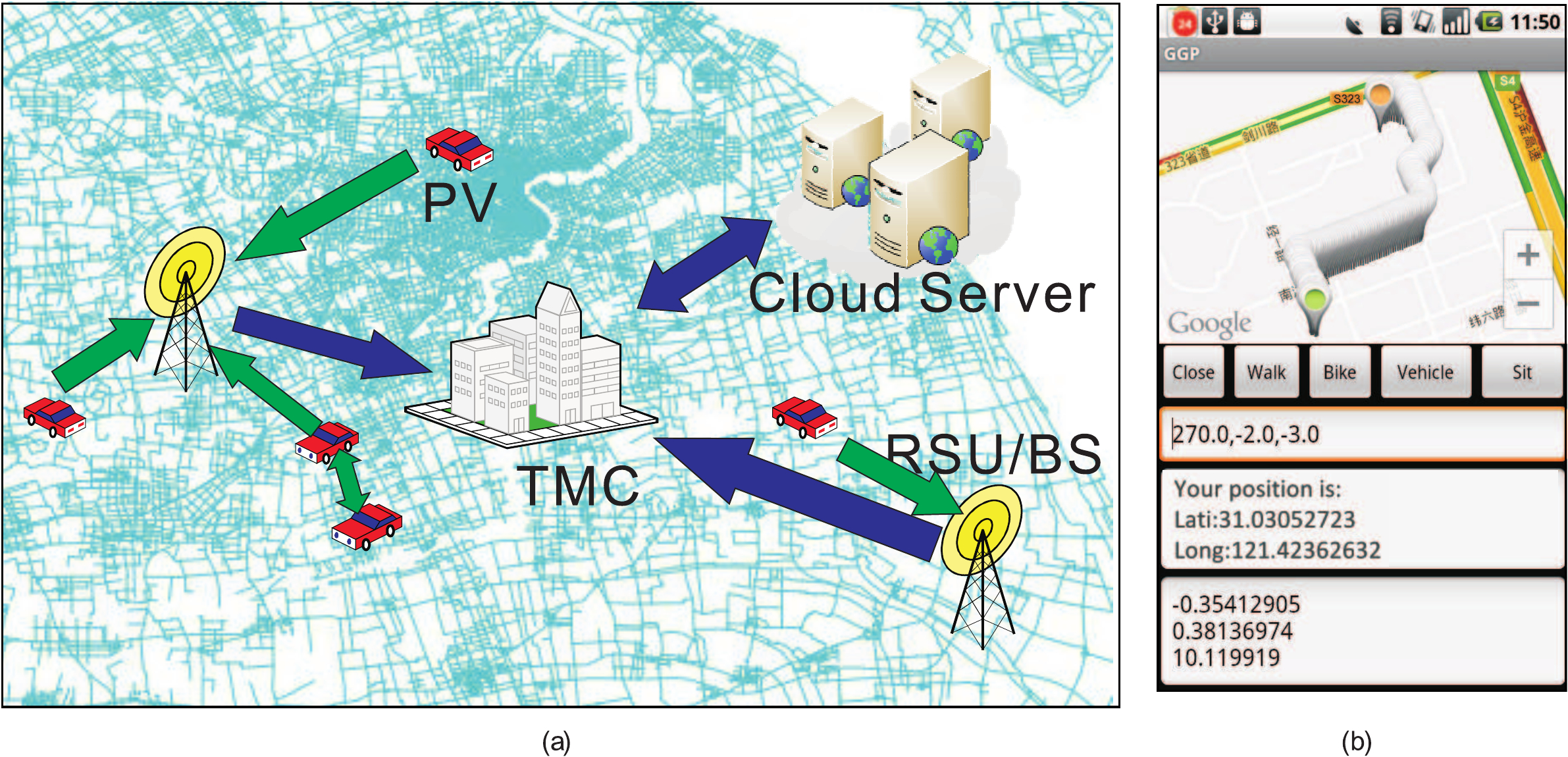}
  \caption{Networking scheme for urban traffic estimation in Shanghai, China (left); The designed Android APP for Automatic Offloading Engine (right)}
  \label{SHscheme}
\end{figure}

In this section, we describe a prototype of TrasoNET based data analytics for realtime traffic sensing and service provisioning. Based on the framework depicted in Fig.~\ref{fig:arch}, the prototyping system consists in three basic components: probe vehicles (PVs), TMC, and a cloud server for traffic analysis and network access recommender. Fig. \ref{SHscheme} shows the system structure which is applied to estimate the traffic of Shanghai, China based on real GPS dataset of $32,122$ taxies (as PVs) on Jan. 24, 2013. To offload the cellular data traffic, the city-wide WLANs have been developed in Shanghai and the number of AP is over $130,000$ including \emph{i-Shanghai} free WiFi in important social spots. Thus the cellular network and WLAN network form the access network layer. The TMC and cloud server are on the data aggregation layer. The application considered in this case is the on-demand network service provisioning for vehicles in a certain region.

\subsection{Data analytics for traffic sensing in CVTS} \label{sec:trafficsensing}
The taxies in Shanghai generate sensing reports every $30$ seconds and report the readings of GPS, i.e. location, time of report, current speed and headings, to TMC through cellular networks. The TMC collects all the traffic reports and constructs a huge traffic matrix $X=\{x_{ij}\}$, in which each entry $x_{ij}$ represents the traffic condition of the $i$-th road (e.g. average speed based on all the reports from PVs on the road) at the $j$-th duty cycle of a day. For example, the location of each report is matched to one road by map matching algorithm, data from different PVs are fused to get the traffic matrix $X$. Since the PVs cannot cover all the roads for all the time, TMC needs to estimate the traffic of un-sampled road in the traffic matrix. Matrix completion is applied in \cite{RDu2015,RDuGlobecom2013} with the low rank property of the traffic matrix. The matrix completion based estimation could be computed in the cloud. The estimation result is then sent to TMC for traffic management and message publishing to the vehicles in the city through the traffic bulletin board or information push through cellular network.

The main idea of the real trace analytics is as follows.
Firstly, estimate the values of average speed in the un-sampled roads \cite{CLChen2014springer} with the constraints of the temporal continuity and bound of the traffic data (the speed limit), respectively.
Secondly, use the sampled data, together with the estimated data, to solve the optimization problem by minimizing the rank of the traffic matrix. The so-called HaTTEM algorithm is presented in \cite{RDuGlobecom2013}.

\begin{figure}[t]
\centering
    \includegraphics[width=4 in]{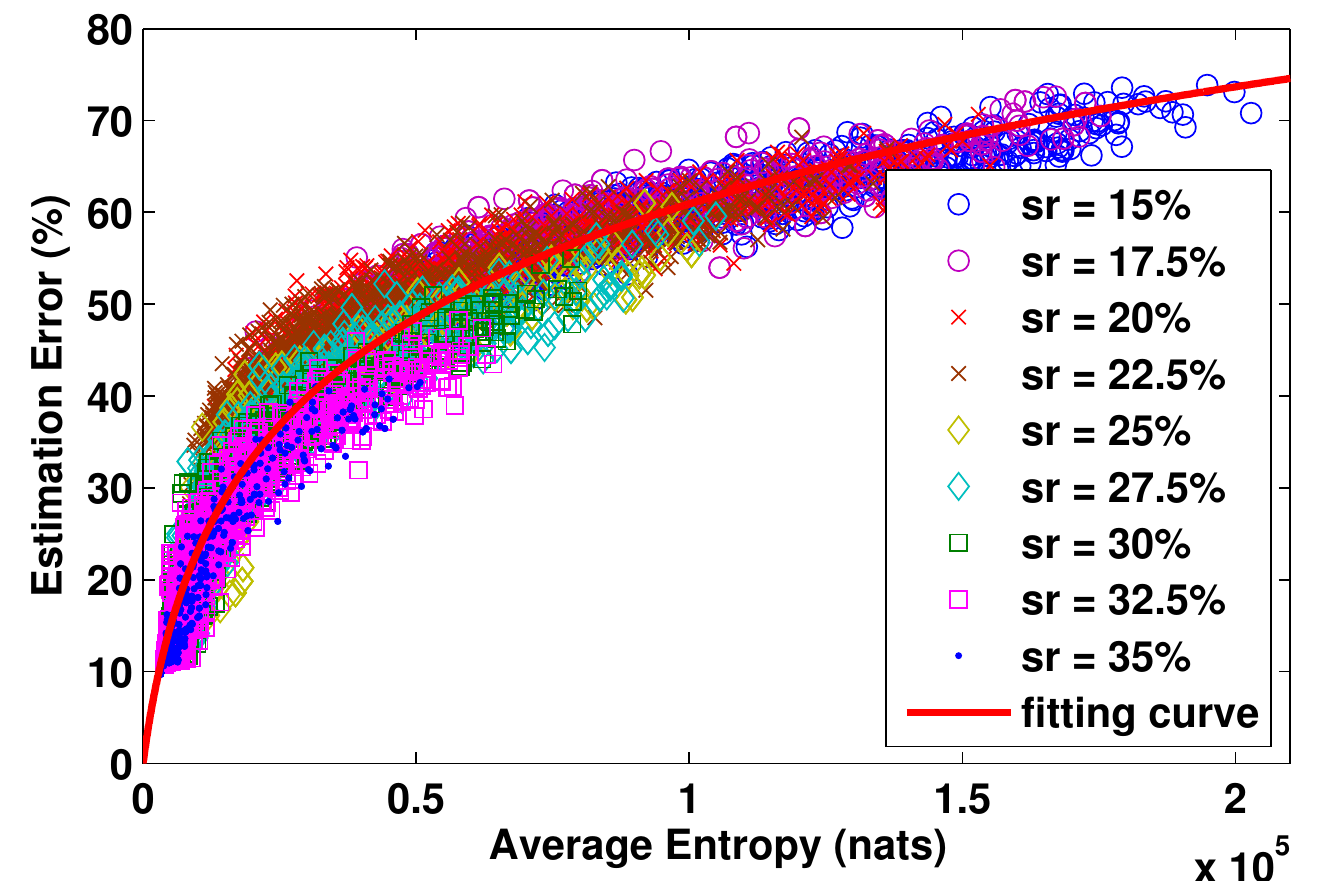}
  \caption{Estimation error VS average entropy under different sample rate}
  \label{fullAverageEntropy}
\end{figure}

However, the integrity analysis of sensing report about specific roads tells the fact that only $69\%$ of over $35,000$ roads in Shanghai have sensing reports for only $30\%$ time of the day. There are not any GPS reports of taxies or buses in 17\% roads within a whole day \cite{RDu2015}. The coverage of taxies' traces in the city is quite uneven due to the aforementioned social proximity. So we need floating cars (FCs, e.g. police cars from Public Security Bureau and patrol cars from Traffic Management Department) to provide more data. These cars don't need to change the patrolling area, but just adjust the patrolling path for better traffic sensing. It is well-known that the disorder of samples can be expressed by entropy. The relation between the entropy and the estimation error is seen in Fig. \ref{fullAverageEntropy}. By planning the paths of only $260$ controllable FCs for the whole city of Shanghai, China, even with the $15\%$ of current PV samples, the average entropy is reduced from $0.233$ nats (unity of entropy) to $0.05$ nats. Thus, it is seen from Fig. \ref{fullAverageEntropy} that the estimation error could be reduced from $35\%$ to $15\%$ with the complimentary GPS data of FCs.

\subsection{Recommendation algorithm for Traffic-dependent networking }

The PVs and FCs connect not only to TMC for management, but also frequently to Internet for providing more emerging services, (e.g., the new free taxi calling services in Shanghai, China with mobile APPs called Diditaxi and Kuaidadi\footnote{URLs: www.xiaojukeji.com; www.kuaidadi.com}). Access to Internet would become a standard feature of future motor vehicles. However, simply using the cellular infrastructure for vehicle Internet access may result in an increasingly severe data overloading issue, which eventually would degrade the communication service performance of both traditional smartphone and vehicular mobile users. This advances of citywide free WLAN access in Shanghai make it possible to serve vehicular users in the near future. This article provides an access network recommendation mechanism for different network applications based on the estimated traffic which could be achieved by the method in Subsection \ref{sec:trafficsensing}.

\begin{figure*}[htbp]
\centering
\includegraphics[width=6in]{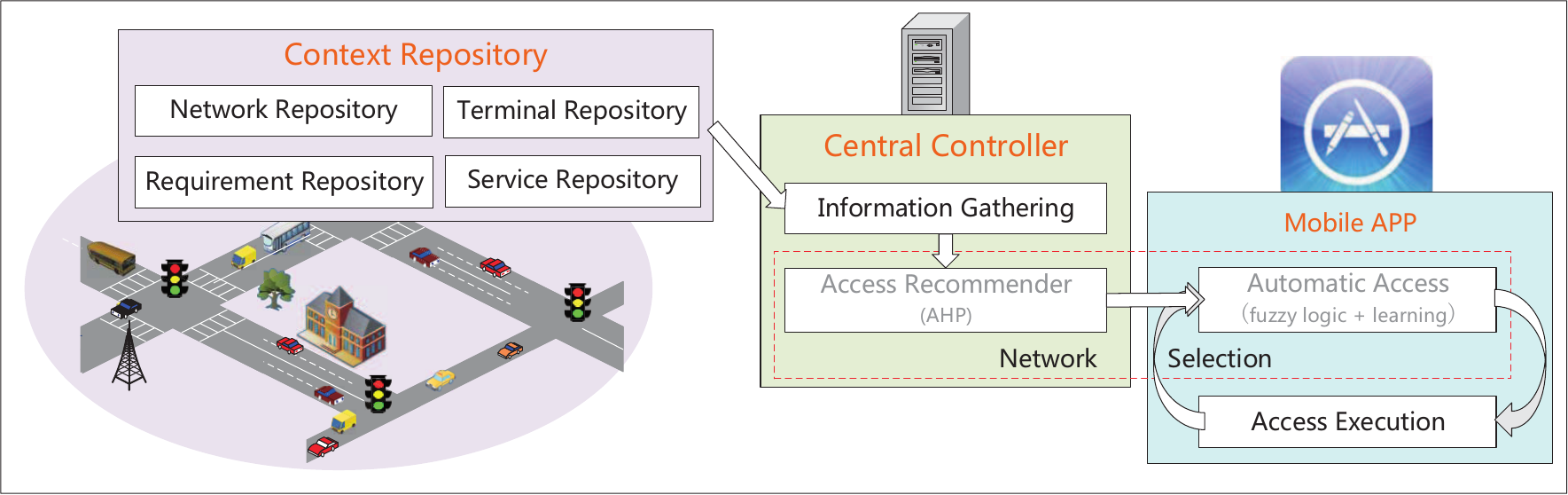}
\caption{Block diagram of intelligent network access system}
  \label{fig:block}
\end{figure*}

\begin{table*}[htbp]
\caption{Comparison matrix based on AHP}
\label{Table1}
\centering
\begin{tabular}{c|c|c|c|c|c||c}
\hline
Service Type & Criterion & Traffic Density & Bandwidth & Delay & Payment & Network Priority \\
\hline\hline
 & Traffic Density & 1 & 5 & 3 & 7 & 0.5558\\
 & Bandwidth & $1/5$ & 1 & $1/3$ & 5 & 0.1364 \\
 & Delay & $1/3$ & 3 & 1 & 5 & 0.2589 \\
\rb{Voice} & Cost & $1/7$ & $1/5$ & $1/5$ & 1 & 0.0489 \\
 \hline
 & Traffic Density & 1 & $1/7$ & $1/5$ & $1/3$ & 0.0553\\
 & Bandwidth & 7 & 1 & 3 & 5 & 0.5650 \\
 & Delay & 5 & $1/3$ & 1 & 3 & 0.2622 \\
\rb{Video}  & Cost & 3 & $1/5$ & $1/3$ & 1 & 0.1175 \\
 \hline
\end{tabular}
\end{table*}

In order to demonstrates the feasibility of traffic-dependent networking, we provide in Fig. \ref{fig:block} the intelligent network access system (INAS) for efficient and economical communication. INAS consists of network recommender by TMC and automatic access engine in mobile devices. It works in three phases, i.e. information gathering, network selection and access execution. The context repository module in Fig. \ref{fig:block} is the knowledge base in Fig. \ref{fig:app}.

\begin{figure*}[htbp]
\centering
  \subfigure[Celluar Network for Voice Service]{
  \includegraphics[width=0.35\textwidth]{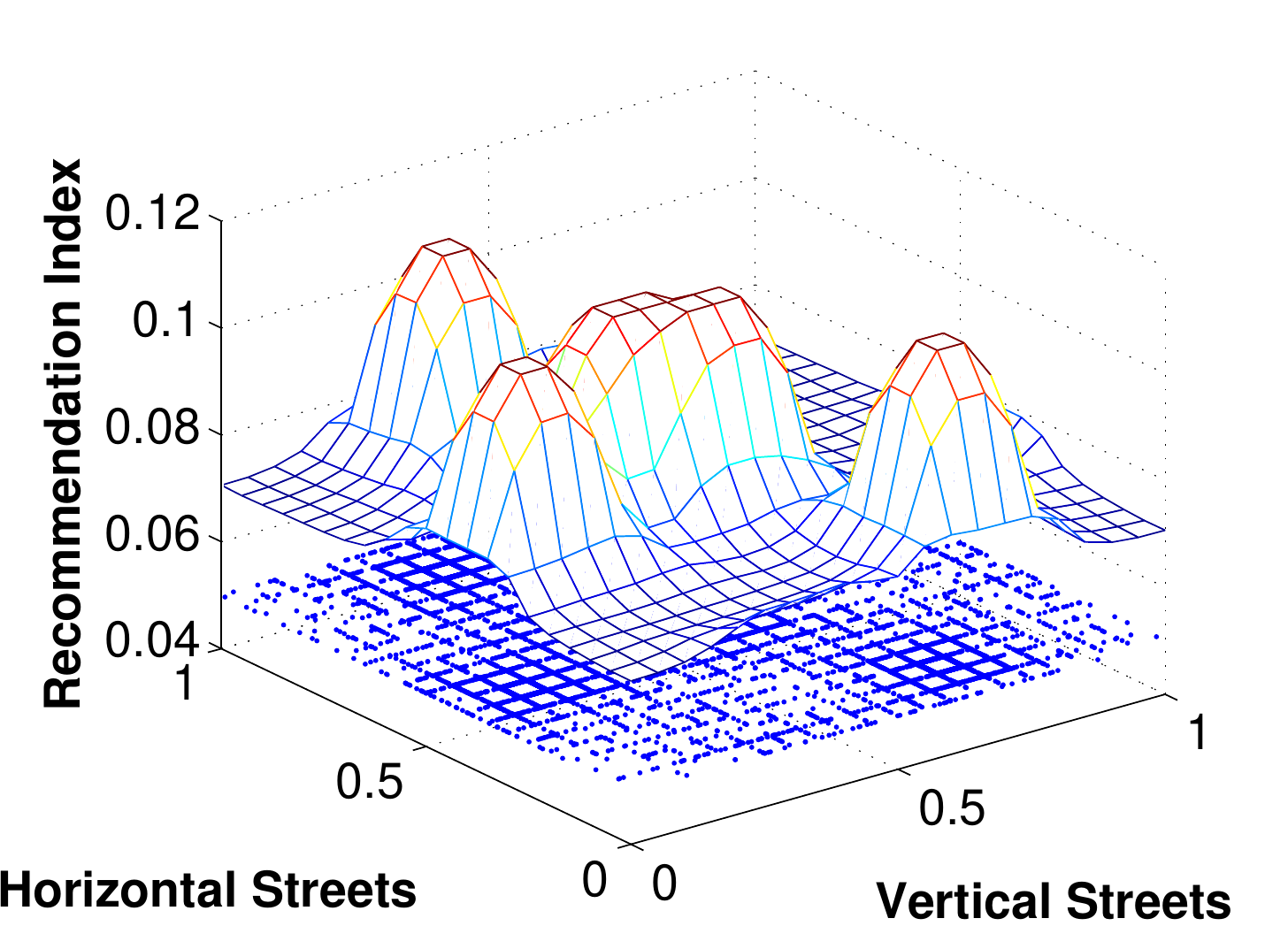}}
  \hspace{1cm}
  \subfigure[VANET for Voice Service]{
  \includegraphics[width=0.35\textwidth]{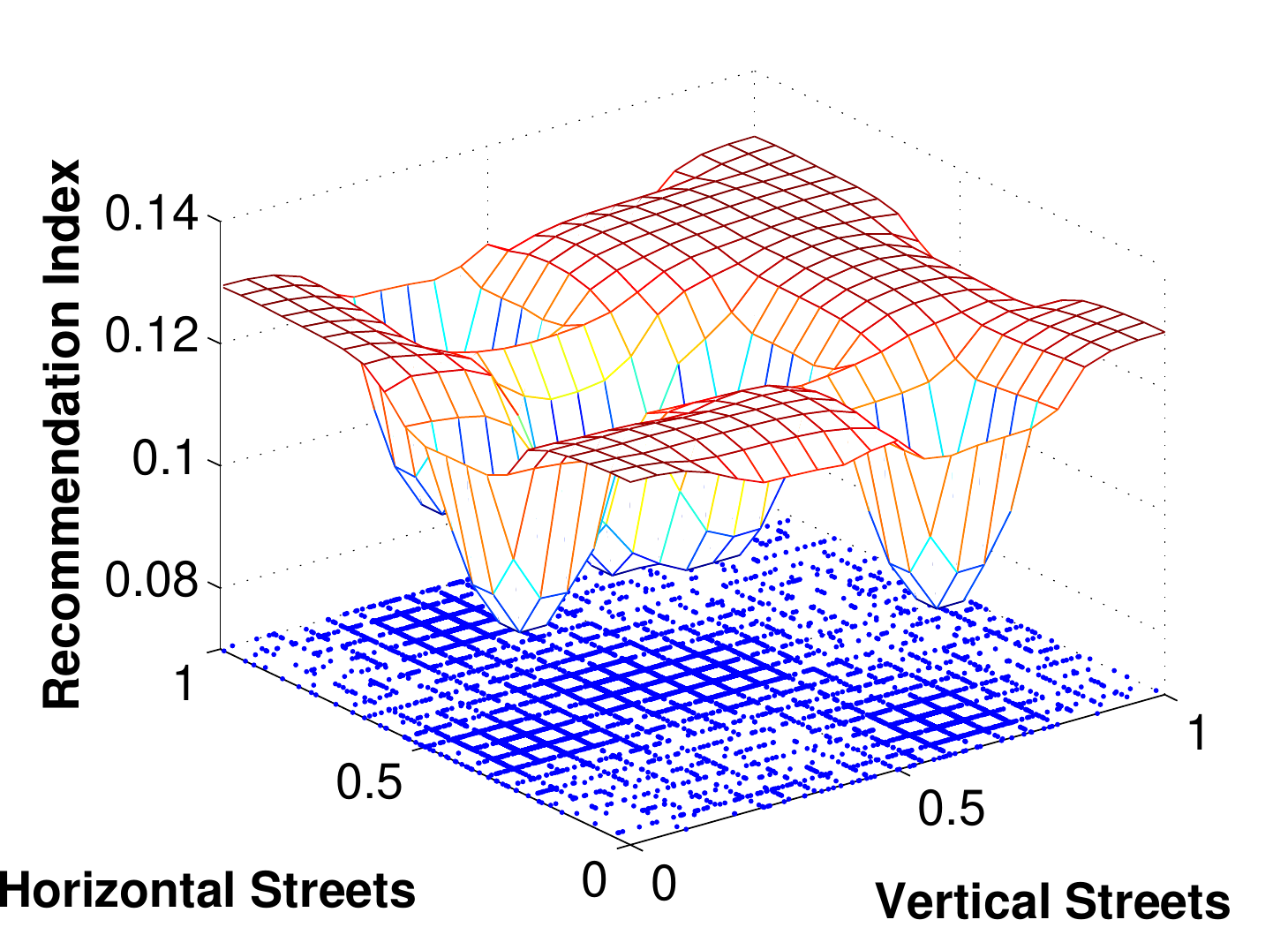}}
  \hspace{1cm}
  \subfigure[Celluar Network for Video Service]{
  \includegraphics[width=0.35\textwidth]{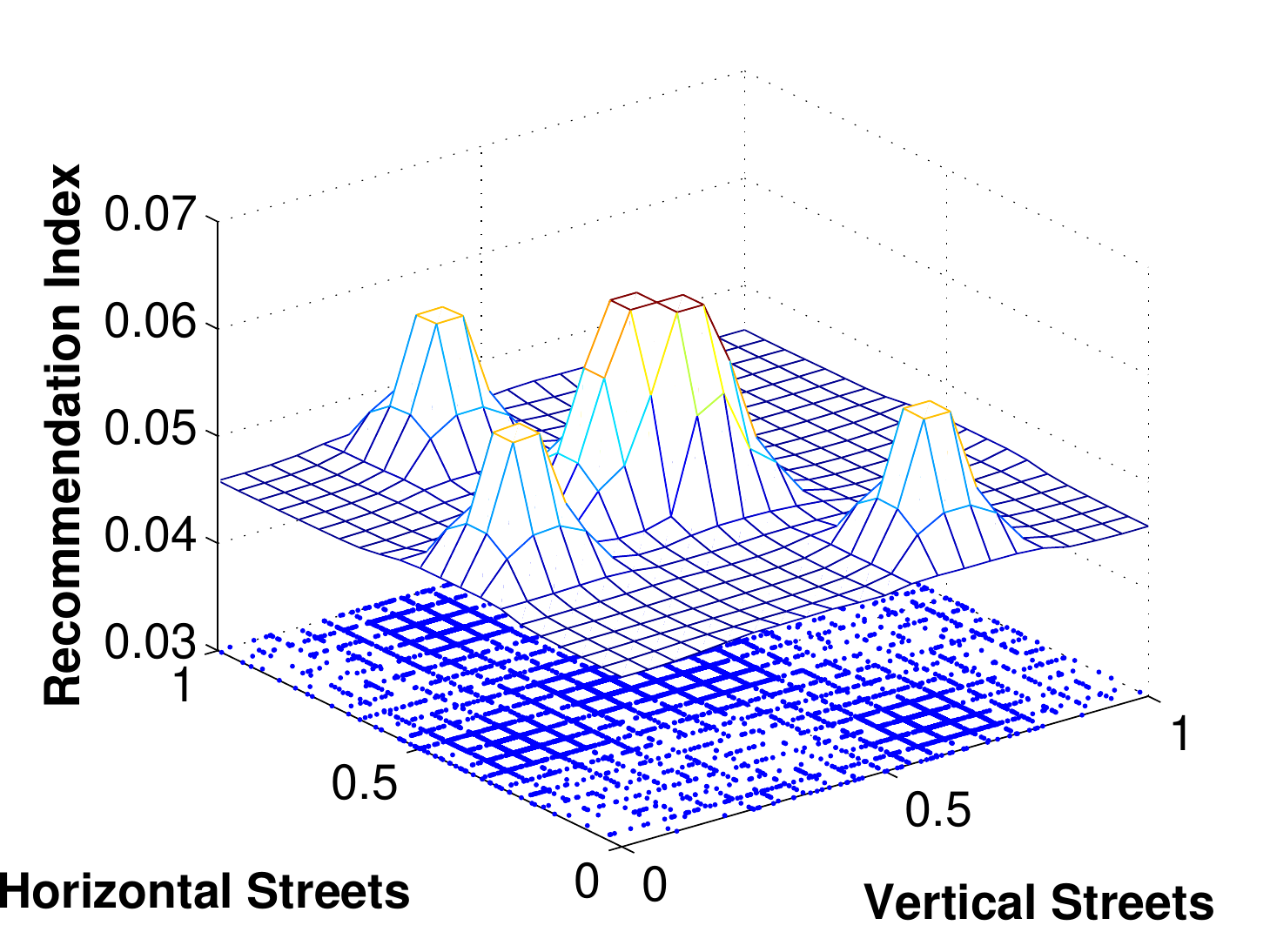}}
  \hspace{1cm}
  \subfigure[VANET for Voideo Service]{
  \includegraphics[width=0.35\textwidth]{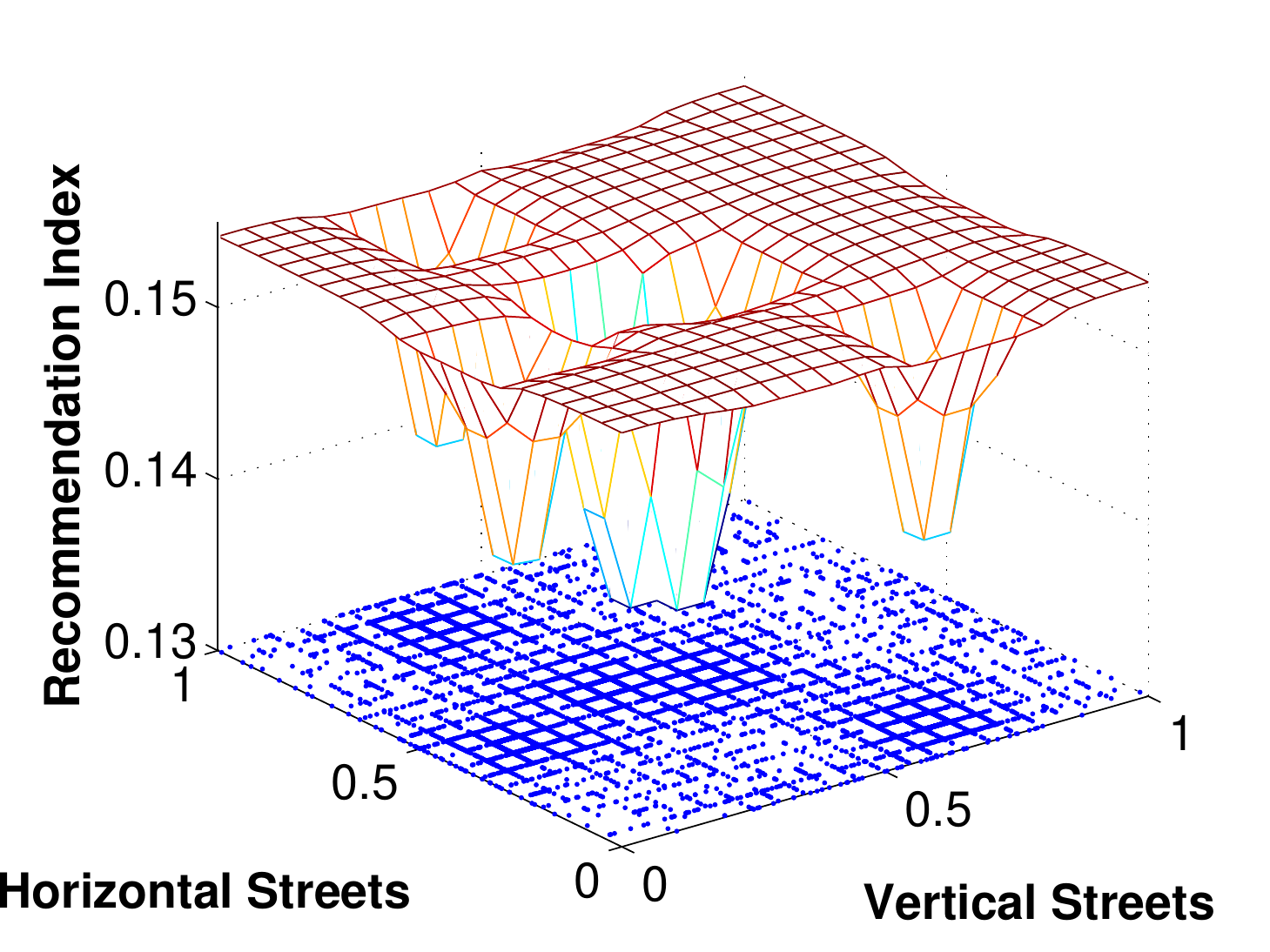}}
  \caption{Recommendation of cellular network and VANET for voice and video services}
  \label{fig:video}
\end{figure*}

Model the urban traffic as scalable grids. In the simulation, consider $5$ SPs and $20,000$ vehicles in the area of $10$KM$\times10$KM with restricted mobility region for each vehicle. There are $20$ vertical and horizontal streets, respectively. Assume that vehicles mobility region is partitioned into multiple tiers co-centered at their SPs. The distribution of mobility follows social proximity model and the vehicle dense obeys the power-law decaying from the center of SP to the border of the mobility region with the exponent $\gamma=2$. Without loss of any generality, consider two types of real-time applications, i.e. Voice Service and Video Service for individual vehicles. Assume the service requirements are $3$ minutes of voice and $5$ minutes of video on average. The data flow rate is $0.6$Kbps and $5$Mbps for voice and video service, respectively. The data rates are RMB$1/$Mb for cellular network and RMB$10/2$Gb per month, respectively. Based on the aforementioned AHP method, the network access recommendation can be given based on the comparison matrices in Table \ref{Table1}. It implies that Voice Service is sensitive to network access delay, while Video Service need more priority for bandwidth. Furthermore, we show the access network recommendation result in Fig. \ref{fig:video} for different service types according to the traffic condition (vehicle density) demonstrated on the bottom X-Y layer of Fig. \ref{fig:video}.

For the traffic density, its second level pair-wise comparison matrix is formed as $A_{2\times2}$. The simulation result about the values of density-tolerance for the two applications (voice and video services) shows that without the network selection algorithm, the successful transmission probability is nearly zero when traffic density is 0.04 for voice service, and 0.06 for video service. The result shows that without the algorithm, it's almost impossible to satisfy every cars' QoS need.

It is noted that Fig. \ref{fig:video} represents the average index values for the recommendation of cellular network and VANET, respectively. If there are only two accessible networks, the priority can be normalized. It is easily seen from Fig. \ref{fig:video} that cellular network is recommended for voice service in a large region around SP where the vehicular traffic density is relatively high. Therefore, cellular network is still the first choice for voice service, especially at SPs. On the other hand, VANET is recommended to offload the cellular network for video service in the region close to SP (except SP due to QoS requirement). It indicates that the network selection/ handover is closely related to vehicular traffic condition, which is demonstrated the necessity of traffic-dependent networking.

With the access network recommendation, the procedure of distributed automatic network access decision-making could be shown in Fig. \ref{fig:app}.There are $4$ premise variables $\langle$$S,A,O,R$$\rangle$, which represent \textbf{S}peed of vehicle, \textbf{A}pplication of network (i.e. voice or vedio), current \textbf{O}ption of access network, and \textbf{R}ecommendation of access network, respectively. The output is achievable QoS represented by $\mathcal{Q}$. The fuzzy sets and corresponding membership functions for each premise variable can be seen in Fig. \ref{fig:app}. $S$ is in the range of $0\sim 80$km/h. The fuzzifier for $A$, $O$ and $R$ is singleton. Hence, we have the following $16$ fuzzy rules:
\begin{itemize}
\item Rule 1: If $S$ is $Low$, $A$ is $Voice$, $O$ is $Cellular$ and $R$ is $Cellular$, then $\mathcal{Q}$ could be $level_{h}$;
\item Rule 2: If $S$ is $Low$, $A$ is $Voice$, $O$ is $Cellular$ and $R$ is $VANET$, then $\mathcal{Q}$ could be $level_{h}$;
\item $\cdots$;
\item Rule 16: If $S$ is $High$, $A$ is $Video$, $O$ is $VANET$ and $R$ is $VANET$, then $\mathcal{Q}$ could be $level_{l}$.
\end{itemize}

With defuzzifier of fuzzy inference result, the distributed automatic network access engine determines the network selection and handover. In order to avoid ping-pong handover due to the mobility and perturbation of QoS, set two thresholds for QoS and delay, respectively. Calculate the QoS improvement by switching the current network to the other. Only if the improvement exceeds the QoS threshold for the time longer than the delay threshold, the handover happens.

\section{Conclusion and Future Research Topics}

This article describes an architecture called TransoNET for data analytics and networking in connected vehicles enabled transportation systems. To efficient manage network resources for CVTS applications, we describe the features of data analytics and subsequently introduce the traffic-dependent networking approach for data collecting. It shows how vehicular traffic can be estimated by matrix completion and how the recommendation-automatic integrated method provides efficient guidance to vehicles for network accessing. The data analysis based on real traces of taxies gives an exemplary study on traffic sensing. In particular, we study a case of multiple-network selection by the combination of network access recommendation from cloud and automatic access engine in vehicles. It has been demonstrated the necessity to explore the relationship between vehicular traffic and networking for providing real-time services in CVTS.

Based on the proposed CVTS architecture, potential research directions can be envisioned to improve the data analytics and networking performance from both cloud and vehicle sides. On the cloud side, big data processing algorithm can be incorporated, e.g. crowdsourcing technologies, for ubiquitous traffic sensing such that more vehicles could take the roles of PV and FC. The social patterns of the vehicles may be considered to improve the traffic crowdsensing. On the vehicle side, automatic network access engine needs low-complexity decision-making algorithms for explosively increasing infotainment services through vehicles to Internet connection. As the terminals of crowdsensing, the vehicles could be more intelligent by automatically adapting the cycles of sensing and reporting according to local vehicular traffic. We believe CVTS will attract enormous attention from academia and industry in the near future.

\bibliographystyle{IEEEtran}
\bibliography{bigdata}

\section*{Biographies}
\vspace{-2cm}
\begin{IEEEbiographynophoto}{Cailian Chen}
(cailianchen@sjtu.edu.cn) is currently a Professor of Shanghai Jiao Tong University, China. Her research interests include vehicular ad hoc networks, wireless sensor and actuator network and computational intelligence. Dr. Chen was one of the First Prize Winners of University Natural Science Award from The Ministry of Education of China in 2007. She received the "IEEE Transactions on Fuzzy Systems Outstanding Paper Award" in 2008. She was honored "New Century Excellent Talents in University" by Ministry of Education of China, "Pujiang Scholar" and "Shanghai Rising-Star" by Science and Technology Commission of Shanghai Municipality, China.
\end{IEEEbiographynophoto}
\vspace{-2cm}
\begin{IEEEbiographynophoto}{Tom Hao Luan}
(tom.luan@deakin.edu.au) recieved the B.Eng. degree from Xi'an Jiao Tong University, China, in 2004, M.Phil. degree from Hong Kong University of Science and Technology in 2007, and PhD degree from University of Waterloo, Canada, in 2012. He is currently a Lecturer in the School of Information Technology at the Deakin University, Melbourne, Australia. From March 2013 to August 2013, he was a visiting research scientist in the Institute of Information Engineering, Chinese Academy of Sciences.
\end{IEEEbiographynophoto}
\vspace{-2cm}
\begin{IEEEbiographynophoto}{Xinping Guan}
(xpguan@sjtu.edu.cn) is currently a Distinguished Professor of Shanghai Jiao Tong University, China. He is also the Professor of "Cheung Kong Scholar" Program, appointed by Ministry of Education of P. R. China, and the winner of "National Outstanding Youth Foundation", granted by NSF of China (NSFC). His current research interests include wireless sensor networks, cognitive radio and wireless technologies for smart grid and smart community. He received First Prize Winners of University Natural Science Award from The Ministry of Education of China in 2006, and the Second Prize of National Natural Science Award from The Ministry of Science and Technology of China in 2008. He received the "IEEE Transaction on Fuzzy Systems Outstanding Paper Award" in 2008.
\end{IEEEbiographynophoto}
\vspace{-2cm}
\begin{IEEEbiographynophoto}{Ning Lu}
(nlu@tru.ca) received the B.Sc. and M.Sc. degrees from Tongji University, Shanghai, China, in 2007 and 2010, respectively, and PhD degree from University of Waterloo, Waterloo, Canada in 2015. He is currently an Assistant Professor in the Department of Computing Science at Thompson Rivers University, Canada. His research interests include capacity and delay analysis, media access control, and routing protocol design for vehicular networks.
\end{IEEEbiographynophoto}
\vspace{-2cm}
\begin{IEEEbiographynophoto}{Yunshu Liu}
(5110309437@sjtu.edu.cn) is pursuing the M.Sc. degree at the Department of Automation, Shanghai Jiao Tong University, China. His research interests include vehicular network based traffic monitoring and application of compressive sensing.
\end{IEEEbiographynophoto}
\end{document}